\documentclass[9pt,twocolumn,twoside]{pnas-new}

\templatetype{pnasresearcharticle}

\usepackage{caption}
\usepackage{subcaption}
\usepackage{ulem} 
\newcommand{\onlinecite}[1]{\hspace{-1 ex} \nocite{#1}\citenum{#1}}

\newcommand{\eb}{\textcolor{black}}

\newcommand{\revise}{\textcolor{black}}

\renewcommand\thesubsection{\arabic{subsection}}
\renewcommand\thesubsubsection{\arabic{subsubsection}}
\makeatletter
    \def\@seccntformat#1{\@ifundefined{#1@cntformat}%
       {\csname the#1\endcsname\space}
       {\csname #1@cntformat\endcsname}}
    \def\subsection@cntformat{\thesection.\thesubsection\space} 
    \def\subsubsection@cntformat{\thesection.\thesubsection.\thesubsubsection\space}
\makeatother

\usepackage{fancyhdr}
\usepackage{blindtext}
\usepackage{nameref}
\DeclareUnicodeCharacter{0301}{\'{e}} 
\DeclareUnicodeCharacter{2009}{\,} 
\DeclareUnicodeCharacter{2212}{-}

\title{Driving and characterizing nucleation of urea and glycine polymorphs in water}

\author[a,$\dagger$]{Ziyue Zou}
\author[b, $\dagger$]{Eric R. Beyerle} 
\author[c]{Sun-Ting Tsai}
\author[a,b,1]{Pratyush Tiwary}

\affil[a]{Department of Chemistry and Biochemistry, University of Maryland, College Park, MD 20742}
\affil[b]{Institute for Physical Science and Technology, University of Maryland, College Park, MD 20742}
\affil[c]{Department of Physics, University of Maryland, College Park, MD 20742}

\leadauthor{Zou and Beyerle} 

\significancestatement{Although a common occurrence experimentally, in many instances, a detailed picture of crystal nucleation remains unresolved. Computer simulations are helpful for gaining molecular insights regarding nucleation, but typically unable to reach experimental timescales where nucleation events are readily observed. To overcome this limitation, we use a machine learning augmented molecular dynamics approach that automatically learns and biases along optimized reaction coordinates to enhance sampling of nucleation. We apply this procedure to aqueous solutions of urea and glycine, sampling multiple polymorphs for both and estimating their solvated free energies  relative to their aqueous solvated states. Our reaction coordinates provide evidence of nonclassical nucleation mechanisms for both systems. This protocol should be to all-atom resolution studies of nucleation in different environments.}

\title{Driving and characterizing nucleation of urea and glycine polymorphs in water}

\authorcontributions{ZZ, EB, ST, PT designed research. ZZ, EB performed research. ZZ, EB, PT wrote the manuscript.}
\authordeclaration{There are no competing interests to declare.}
\equalauthors{\textsuperscript{$\dagger$}Ziyue Zou contributed equally to this work with Eric R. Beyerle.}
\correspondingauthor{\textsuperscript{1}To whom correspondence should be addressed. E-mail: ptiwary@umd.edu}


\begin{abstract}
\label{sec:abstract}
Crystal nucleation is relevant across the domains of fundamental and applied sciences. However, in many cases its mechanism remains unclear due to a lack of temporal or spatial resolution. To gain insights to the molecular details of nucleation, some form of molecular dynamics simulations is typically performed; these simulations, in turn, are limited by their ability to run long enough to sample the nucleation event thoroughly. To overcome the timescale limits in typical molecular dynamics simulations in a manner free of prior human bias, here we employ the machine learning augmented molecular dynamics framework ``Reweighted Autoencoded Variational Bayes for enhanced sampling (RAVE)". We study two molecular systems, urea and glycine in explicit all-atom water, due to their enrichment in polymorphic structures and common utility in commercial applications. From our simulations, we observe \revise{multiple} back-and-forth liquid-solid transitions of different polymorphs \revise{and from these trajectories calculate the polymorph stability relative to the dissolved liquid state. We further observe that the obtained reaction coordinates and transitions are highly non-classical.}

\end{abstract}
\begin{document}

\maketitle
\thispagestyle{firststyle}
\ifthenelse{\boolean{shortarticle}}{\ifthenelse{\boolean{singlecolumn}}{\abscontentformatted}{\abscontent}}{}

\section*{Introduction}
\label{sec:introduction}

Although traditionally viewed through the lens of classical nucleation theory (CNT) \cite{Peters2017}, which postulates the only important variable for describing nucleation and subsequent crystal growth is the size of the nascent crystal \cite{Karthika2016}, more recent studies of nucleation for a number of physical systems under a variety of conditions via high-resolution experiments \cite{Nielsen2014,Nakamuro2021,deYoreo2015,Abecassis2022persistent} and molecular dynamics (MD) simulations \cite{piaggi2022icenucleation,LaCour2022Tuning,bertolazzo2022zeolite,Jacobson2010,Shtukenberg2019paracetamol,Karthika2016,Giberti2013, Finney2022naclnucleation, tsai2019reaction,Salvalaglio2012uncovering, Niu2019IceNucleation,gobbo2018infoS} have shown that CNT alone is not sufficient for an accurate description of the mechanism of crystal formation, which often involves multiple, competing pathways \cite{deyoreo2013morethan, deYoreo2020, Nielsen2014}. Thus it is of interest to discover other important variables besides or in addition to the size of the nucleus to accurately describe the nucleation process. One can imagine this situation is especially relevant for substances such as \revise{H$_2$O} or organic molecules possessing multiple stable, crystalline structures or polymorphs, where both the lattice structure as well as crystal size are important in distinguishing the solid and the liquid states from each other.

From an experimental perspective, nucleation is a ubiquitous phenomenon, but the microscopic processes driving the nucleation event typically occur over time- and lengthscales that are, generally speaking, below the resolution of current physical apparatuses \cite{Nakamuro2021,Nielsen2014}. Thus, applying a combination of theory and simulation to describe the microscopic nature of nucleation is required to gain a detailed understanding of the molecular nature of the nucleation process and build a bridge to the time- and lengthscales accessible by computational and experimental approaches.


Even with supercomputers capable of simulating hundreds of microseconds per day \cite{Shaw2021}, the timescale for sampling most nucleation processes, occurring on the timescale of seconds, still remains out-of-reach. Since nucleation is a rare event from the microscopic perspective, nucleation events can generally only be observed in unbiased simulations under conditions of heavy supersaturation or by applying an external driving force. Thus, finding excellent reaction coordinates (RCs) along which the system can be biased to accelerate the sampling is still very much of importance through a variety of enhanced sampling methods \cite{Torrie1977US, laio2002escaping,PhysRevLett2005FFS,JCP2006FFS,rosales2020seeding,bazterra2002modified}. In parallel to the development of more powerful computing machines and sampling algorithms, there has also been an explosion in the use of machine learning (ML) techniques to sift and interpret the results of long simulations \cite{wang2020machine, Ceriotti2019, Hoffmann2021,ghorbani2022gvampnet,banik2022cegan,hong2021MLRC, Mori2020, Rogal2019}.  Neural networks and ML in general have been implemented to discover good RCs for describing nucleation in a more automatic manner. These neural network derived RCs can then be used as the biasing coordinates in different enhanced sampling methods. \eb{In fact, even traditionally RC-free methods, such as path sampling methods (e.g. forward flux sampling) have been shown to benefit from a judicious choice of low-dimensional projections along which to calculate the flux \cite{Defever2019cFFS, VelezVega2010Kinetics}. }

In principle, the RC can be directly expressed as a function of simple variables such as high-dimensional atomic positions. But, for various computational reasons, generally the RC is expressed as a linear or non-linear function of lower-dimensional order parameters (OPs), which are able to distinguish metastable states. These OPs themselves are generally \textit{a priori} designed functions of the atomic positions. Approximating the true RC with smoothly differentiable functions of the OPs has become one of the main aspects of studying crystal nucleation using enhanced sampling approaches such as well-tempered metadynamics (WTmetaD) \cite{PRL2008WTMetaD}, which is our interest in this work, but also more generally for other sampling approaches \cite{giberti2015crystalnucleation,karmakar2021cv,sosso2016crystal,Hussain2020FFSreview,blow2021seven,Peters2006,Ma2005,Borrero2007,zou2021sgoopurea, Schwantes2013, Borrero2007a, DeFever2017, Zimmermann2015nacl, tsai2019reaction, Finney2022naclnucleation, Arjun2021TPS-CO2,Jacobson2011,Badin2021Nacl, Samanta2014, Song2020, Rogal2019}.

In this manuscript, we utilize a neural network framework termed the state predictive information bottleneck (SPIB) \cite{Wang2021SPIB, Shams2022lipidSpib} to extract a two-dimensional RC for describing the nucleation of different polymorphs of urea and glycine from solution. 
 \eb{SPIB is a variant of RAVE that discovers a set of low-dimensional RCs capable of} faithfully predicting the metastable states given the input OPs, which can then be biased in enhanced sampling simulations \cite{Shams2022lipidSpib}. The input OPs originate from a library of candidate OPs and are system specific \cite{Wang2021SPIB, Shams2022lipidSpib}. For the nucleation processes studied here, the OPs are derived from a set reporting on the global and local molecular orientations and packing of the nucleating species, e.g. coordination numbers \cite{tenWolde1998op, tsai2019reaction}, radial distribution functions \cite{Piaggi2017entropy,piaggi2018predicting} and Steinhardt bond OPs \cite{steinhardt1983op}, \textit{inter alia}. We describe SPIB in more detail in~\hyperref[sec:spib]{\textbf{State Predictive Information Bottleneck}}.

We find that biasing MD simulations of urea and glycine in water using the RCs discovered by SPIB allows for a sufficient enhancement of the sampling such that multiple back-and-forth liquid to \revise{solvated} crystal polymorph transitions are observed over practical compute times on high-performance resources. Furthermore, through the use of a linear architecture during the encoding of the RCs, we are able to directly interpret the RCs discovered. Finally, by reweighting these biased simulations \cite{tiwary2015jpcb} the relative stability among \revise{the solvated} crystalline phases sampled is ranked for both urea and glycine, \revise{and these rankings are compared with previous, related studies}.

\section*{Discovering and Biasing Reaction Coordinates for Nucleation}
  \label{sec:method-maintext}

\subsection*{Order parameters for nucleation}
  \label{sec:op}

As mentioned in the Introduction, in this work as well as in work by others \cite{giberti2015crystalnucleation,karmakar2021cv,sosso2016crystal,Hussain2020FFSreview,blow2021seven,Peters2006,Ma2005,Borrero2007,zou2021sgoopurea, Schwantes2013, Borrero2007a, DeFever2017, Zimmermann2015nacl, tsai2019reaction, Finney2022naclnucleation, Arjun2021TPS-CO2}, it is common to build the reaction coordinate as a function of OPs that collectively distinguish among competing metastable states, which has led to the development of several rich OPs for the study of nucleation and phase transitions in general:

\begin{enumerate}[leftmargin=*]

\item \textit{Coordination numbers and associated moments}: 
As introduced above, CNT is based on the size of the nucleus, and the formation of such crystal nuclei is closely related to the formation of a denser phase. A continuous and differentiable coordination number of any molecule $i$ can be defined as:
\begin{align}
&c{(i)} = \sum_{j}\frac{1-(r_{ij}/r_c)^6}{1-(r_{ij}/r_c)^{12}}
\label{eq:cnOP}
\end{align}
where $r_c$ is a radial cutoff and $r_{ij}$ denotes the distance between reference sites for two molecules $i$ and $j$. When averaged over all molecules in the system, the coordination number can serve as an approximation to the true reaction coordinate in the study of gas-liquid transitions \cite{Salvalaglio2016argon,tsai2019reaction,Bal2021GeometricFE}.

Inspired by this definition, in this manuscript we specifically consider two sets of populations of molecules, with coordination numbers greater than 8 and 11, denoted as $N_{8^+}$, $N_{11^+}$, respectively. We also consider the second moment of all coordination numbers as an OP, denoted $\mu^2_{c}$ \cite{Salvalaglio2016argon,tsai2019reaction, tenWolde1998op}. However, such coordination number based descriptors are not sufficient to describe the reaction coordinate in events like nucleation from the melt and polymorphism in molecular systems because nucleation is accompanied with the appearance of a space group. Therefore, in addition to the OPs $N_{8^+}$, $N_{11^+}$ and $\mu^2_{c}$ derived from coordination number, we include other OPs into the dictionary, as introduced next.

\item \textit{Steinhardt bond OPs}: These OPs, originally introduced in Ref.~\onlinecite{steinhardt1983op}, have been used to distinguish Lennard-Jones solids \cite{Moroni2005interplay, tenwolde1996numerical, tenwolde1999homogeneous, Volkov2002MDhardsphere, Auer2001LJparticle,Chopra2006LJ,Desgranges2006LJ} and have been extended to the study of ice polymorphs \cite{Radhakrishnan2003iceJACS, Radhakrishnan2003icePRL, Quigley2008ice, Brukhno2008ice, Reinhardt2012ice}. They map a given local environment onto specific degrees of spherical harmonics \cite{steinhardt1983op}, defined through :
\begin{align}
&q_{lm}{(i)} = \frac{\sum_{j} {\sigma(r_{ij})\bf{Y}_{lm}(\bf{r}_{ij})}}{\sum_{j}{\sigma(r_{ij})}}
\label{eq:SteinhardtOP}
\end{align}

where $\bf{Y}_{lm}$ is the $l^{th}$ order of spherical harmonics and $m$ ranges from $-l$ to $l$. As these bond OPs refer to distributions as superpositions of perfect crystalline structure, reliable results have been acquired in classifying face-centered-cubic (fcc), body-centered-cubic (bcc), hexagonal-close-packed (hcp) and liquid-like structures \cite{Moroni2005interplay}. Here, we add $\bar{q}_4$, $\bar{q}_6$ from Ref.~\onlinecite{steinhardt1983op} corresponding to the average values of the fourth and sixth orders of spherical harmonics to our OP library. However, as pointed out by Dellago and Lechner \cite{lechner2008accurate} thermal fluctuations may destroy the ability in phase identification and, as such, the variants of original bond OPs have been developed for better differentiating ability at the expense of a higher computational cost; these are omitted here.

\item \textit{Interfacial water}:
The contribution of solvent molecules can hardly be neglected when studying the nucleation of solvated systems \cite{Salvalaglio2012uncovering}. Here we introduce a new OP measuring approximately the population of interfacial waters or more generally solvent molecules surrounding solid-like solute molecules by using tunable distance cutoffs. This variable is defined as 

\begin{align}
    N_s = \sum_i{\frac{1}{2}\xi (i) c_{solvent}(i)}
    \label{eq:IntWatOP1}
\end{align}

where 
\begin{align}
    \xi (i) = \tanh(c_{solute}(i)-c_o)+1
    \label{eq:IntWatOP2}
\end{align}

Here $i$ denotes a sum over solute molecules and $c_{o}$ is a tunable parameter for the determination of solid-like aggregates which is set to be 5.0 for all systems in this work \cite{tenWolde1998op}. Effectively, the OP in Eq.~\ref{eq:IntWatOP1} is a product of two smoothened Heaviside functions: $\xi(i)$ only counts solute clusters larger than the threshold $c_0$, and $c_{solvent}(i)$ is a coordination number between such selected solute atoms and the solvent, as given in Eq.~\ref{eq:cnOP}. Thus, $N_s$ is large when two conditions are simultaneously satisfied: when the solute cluster is sufficiently large and also well-coordinated with solvent. If either one of these conditions is not satisfied, $N_s$ is small. Thus, large values of $N_s$ indicate cluster formation and low values indicate a liquid state or the presence of small, solvent-separated clusters. 

\item \textit{Averaged intermolecular angles and associated moments}: 
The estimation of the size of solid-like aggregates and their solvation states is useful and intuitive for the study of molecular systems in different environments \cite{Salvalaglio2016argon,tsai2019reaction,Bal2021GeometricFE}, but nucleation mechanisms can easily consist of non-classical behavior and polymorphs such that solely distance based OPs fail in categorizing them. In the spirit of Steinhardt bond OPs \cite{steinhardt1983op} and environment similarity OPs \cite{EnvSim2019Piaggi}, we introduce a set of OPs which account for the orientation of molecular feature vectors of solutes to help better discriminate polymorphs. For any molecule $i$, we define:
\begin{align}
    \theta{(i)}=\frac{\sum_{j}{ \sigma (r_{ij})\frac{1}{2}[(\pi-2\tilde{\theta})\tanh(5\tilde{\theta}-9.25)+\pi]}}
    {\sum_{j}{\sigma (r_{ij})}}
    \label{eq:thetaOP}
\end{align}
where $\tilde{\theta}$ is the angle formed between characteristic vector on molecule $i$ and molecule $j$ as illustrated in Fig.~\ref{fig:ang-def}, and $\sigma$ is a switching function of intermolecular distance. The hyperbolic tangent switching function is applied to remove the mirror image symmetry (see SI for detailed explanation). We then compute the mean of angular distribution functions, denoted as $\bar\theta_1$ and $\bar\theta_2$ where 1 and 2 denote the two intramolecular vectors specified for crystalline configuration identification (see Fig.~\ref{fig:ang-def} (a) and (b) for definitions of selected vectors for urea and glycine molecules, respectively). In addition to these OPs, we also consider in our OP dictionary the second moments $\mu^2_{\theta_1}$ and $\mu^2_{\theta_2}$ of the collection of $\theta$ values defined through Eq.~\ref{eq:thetaOP}.

\begin{figure}
  \centering
  \includegraphics[width=8cm]{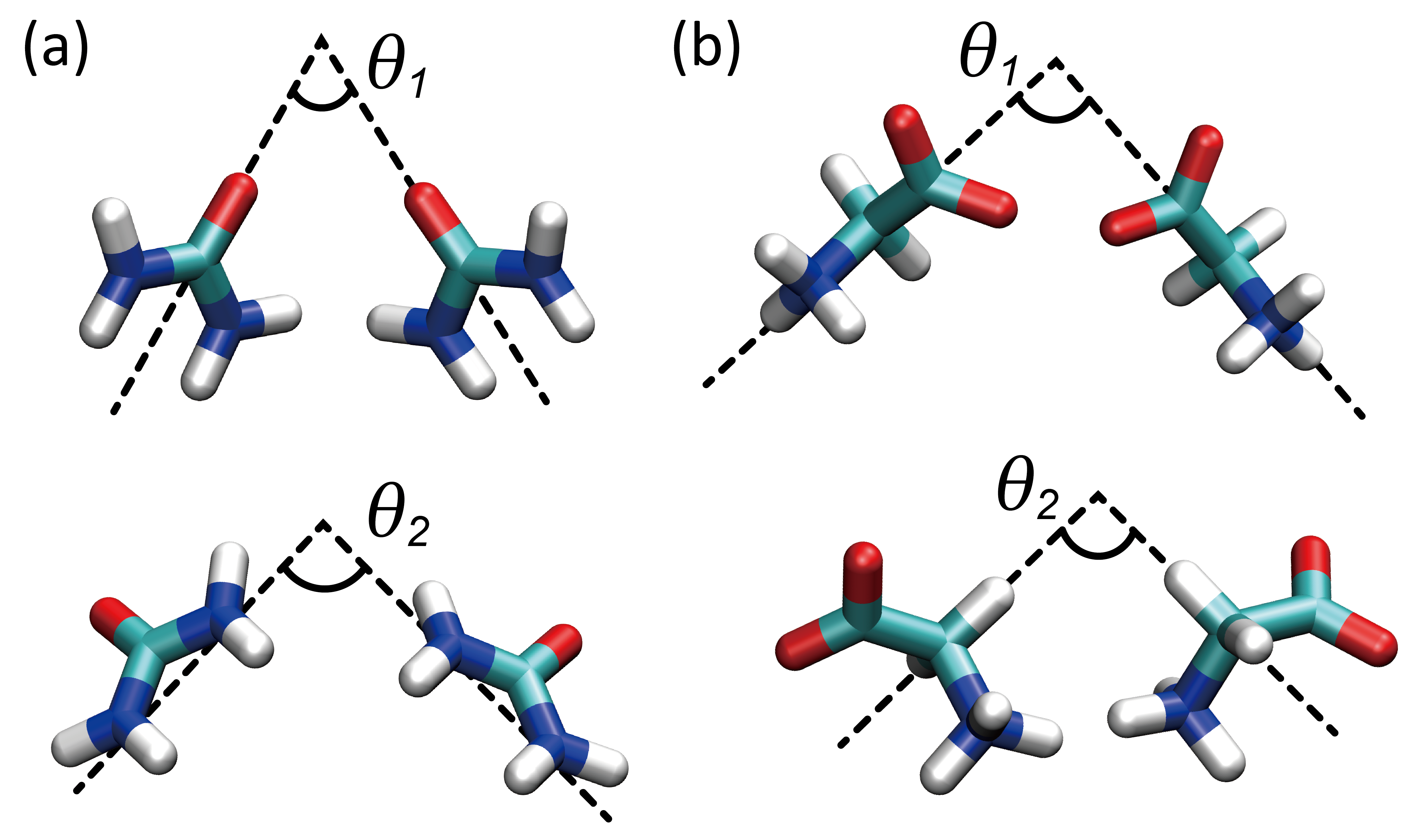}
  \caption
  {Visualization of the feature vectors and corresponding intermolecular angles $\theta_1$ and $\theta_2$ for (a) urea and (b) glycine molecules. The feature vectors shown here are the vectors connecting C-O and N-N for urea, and C-N and C$_\alpha$-H$_\alpha$ for glycine. Different atom types are indicated with different colors, specifically carbons in cyan, oxygens in red, nitrogens in blue, and hydrogens in white. Snapshots are rendered using Visual Molecular Dynamics (VMD) \cite{HUMPHREY1996VMD}. 
  }
  \label{fig:ang-def}
\end{figure}

\item \textit{Pair entropy}: In Ref.~\onlinecite{Piaggi2017entropy} Piaggi and co-workers introduced an OP which approximates radial and orientational entropies from radial different distribution functions. This OP is formulated as: 

\begin{align}
S(r) = -2 \pi \rho k_{B} &\int_{0}^\infty[g(r)\ln{g(r)}-g(r)+1]r^2dr
  \label{eq:S_r}
\end{align}

where $g(r)$ is the radial distribution function, $\rho$ is the density, and $k_B$ is Boltzmann's constant. This OP was introduced in order to account for the possible entropy-favored crystalline structures \cite{piaggi2018predicting}. An advanced version of this OP is developed in Ref.~\onlinecite{piaggi2018predicting}. This pair orientational entropy OP, denoted $S(r, \theta)$, depends on both the intermolecular distance and  the relative angle $\theta$ between a given intramolecular vector equivalently defined for each solute molecule in the simulation:
\begin{align}
S(r, \theta) = - \pi \rho k_{B} \int_{0}^\infty \int_{0}^\pi &[g(r, \theta)\ln{g(r, \theta)}-g(r,\theta)+1] \notag \\ 
&\times r^2\sin(\theta)drd\theta 
  \label{eq:S_r_theta}
\end{align}

It was found \cite{piaggi2018predicting} that such an approximation to the entropy is useful for distinguishing polymorphs in simulations of urea and naphthalene\revise{, and a variant of Eq. \ref{eq:S_r_theta} is used in \cite{Song2020} to successfully discover and distinguish polymorphs of 1:1 mixtures of resorcinol and urea.} Since Eq.~\ref{eq:S_r_theta} only accounts for a single angle, and not the three angles necessary to specify exactly the relative orientation between molecules, this expansion of the entropy, though useful, is necessarily approximate, as are the other OPs previously described in this section. 
\end{enumerate}

While individually these different OPs might have limitations, they can supplement each other and provide a palette through which the complex molecular processes governing nucleation can be described. It is by combining them through the State Predictive Information Bottleneck (SPIB) \cite{Wang2021SPIB} approach that we construct low-dimensional projections that are much closer to the true RC than any of the OPs on their own. We next briefly summarize the SPIB approach, referring to Ref.~\onlinecite{Wang2021SPIB} for further details.

\begin{table}[b]
    \centering
    \caption{Parameters for WTmetaD}
    \begin{tabular}{|c|c|c|c|c|c|}
    \hline 
        System & $\omega$(kJ/mol) & $\gamma$  & $\sigma_1$ (RC unit) & $\sigma_2$ (RC unit) & $T$ (K) \\ 
          \hline 
         Urea & 5.0 & 100 & 0.2 & 0.2 & 300  \\
         \hline
         Glycine & 5.0 & 100 & 1.15 & 1.15 & 300 \\ 
         
         \hline 
    \end{tabular}
    \label{tab:metad_parameter}
\end{table}

\begin{figure*}[b]
  \centering
  \includegraphics[width=1.0\textwidth]{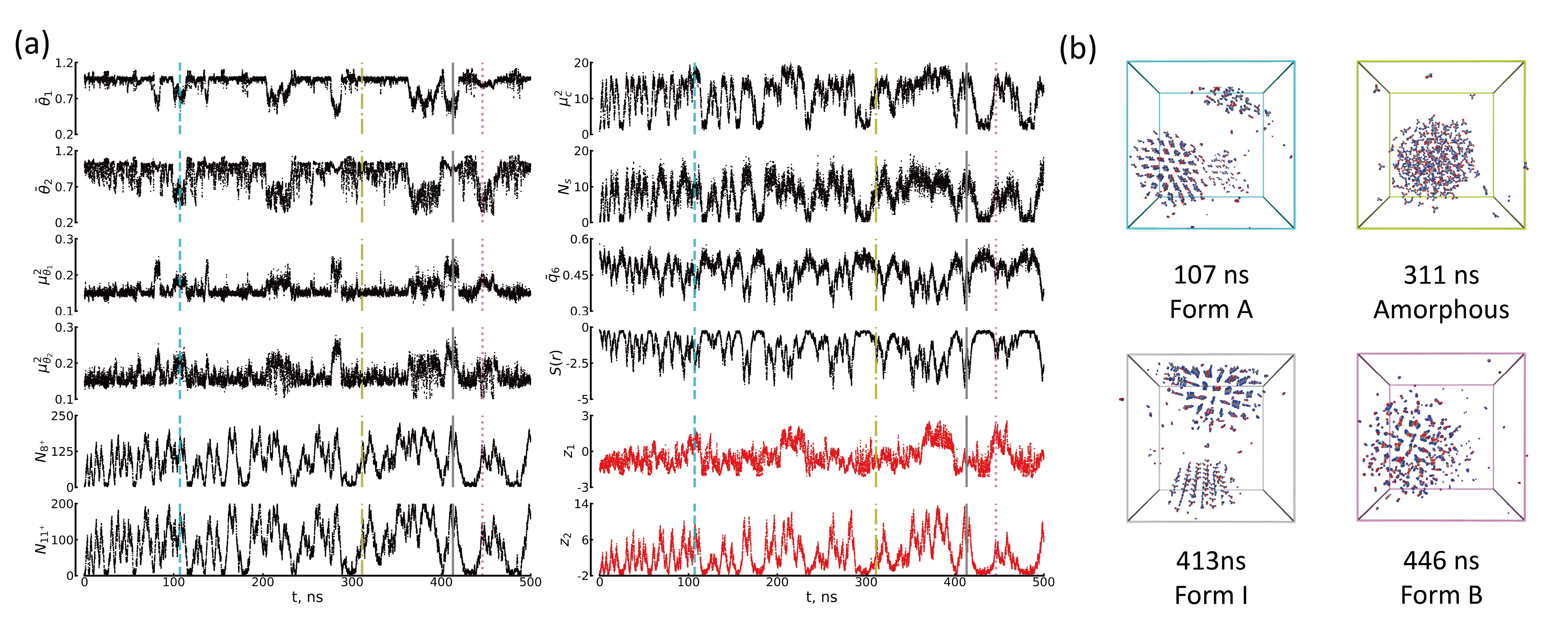}
  \caption
  {Sampling different urea polymorphs with WTmetaD simulations biasing the SPIB-learned 2-dimensional reaction coordinate. (a) shows a representative time series of OPs (black) and reaction coordinates (red), clearly demonstrating multiple back-and-forth transitions between different phases. Vertical cyan dashed, green dash-dotted, grey solid, and pink dotted lines indicate representative transitions to form A, amorphous, form I, and form B from initial liquid phase. (b) shows snapshots of structures sampled during trajectory shown in (a), along with the time (vertical lines in (a)) at which they were observed. Ovito 3.3.5 \cite{Stukowski2010ovito} was the visualization tool used for snapshot generation. 
  }
  \label{fig:urea-sum-md}
\end{figure*}
\subsection*{State Predictive Information Bottleneck (SPIB)}
 \label{sec:spib}
Given the evidence that CNT is insufficient to accurately describe the nucleation event for a number of chemical systems, for instance because knowledge of more than just the solute coordination number OP is required, we need an approach to combine the different possible OPs to develop a lower-dimensional reaction coordinate for describing nucleation in urea and glycine.

Here we utilize a machine learning method known as the state predictive information bottleneck (SPIB) \cite{Wang2021SPIB, Shams2022lipidSpib, Beyerle2022}, which takes the form of a variational autoencoder (VAE) \cite{Kingma2014, Higgins2017}. SPIB differs from the traditional VAE because instead of attempting to maximize the model's ability to re-construct the input from the low-dimensional latent space, it instead maximizes the quality of reproducing the population of the system's metastable states after a given lagtime $\tau$ in the future. That is, the SPIB finds the latent space that best reproduces the metastable state populations at time $t + \tau$ given the values of the input OPs at time $t$ \cite{Wang2021SPIB}. This approach allows SPIB to simultaneously perform dimensionality reduction and accurate future state prediction by minimizing a loss function that is inspired by the variational information bottleneck formalism \cite{Wang2021SPIB, alemi2016deepVIB, tishby2000ib}.

Conceptually, this approach is similar to performing a Markov state model and clustering analysis \cite{Noe2007,Deuflhard2000} with a continuous basis set. In SPIB the number and neighborhood of each metastable state is adjusted on-the-fly during the training of the model. The final learned model is output once the neighborhoods spanned by each metastable state have converged to below a pre-defined threshold that is a tunable hyperparameter of the model.

Finally, since biased simulations are input to the SPIB, normally the WTmetaD weights would be used in the analysis to account for the sampling from a biased distribution \cite{Shams2022lipidSpib}. For urea and glycine, we find that the barriers to nucleation from the liquid state are so large ($\sim$100 kJ/mol or more) that using the metadynamics weights in the analysis effectively destroys any polymorph minima on the surface, yielding a single, global minimum corresponding to the liquid state (more details in the SI). Since SPIB finds the RC describing transitions between metastable minima, the metadynamics weights must be neglected in this case to find an effective set of RCs for nucleation.

\subsection*{Estimating free energies with Metadynamics}
\label{sec:metad}
In order to learn the SPIB, or the approximate RC, as a combination of different OPs, we need access to a trajectory that has visited different possible conformations. Such a trajectory is generated here through the method Metadynamics, which helps the system escape free energy minima by periodically depositing Gaussian kernels as a function of the variable being biased. For the initial metadynamics runs we bias orientationally informative OPs such as intermolecular angles $\bar\theta$ and orientational entropies  $S_\theta$. After performing SPIB analysis on these runs, we then bias along the SPIB-learned RCs in future iterations. In particular, here WTmetaD is used where kernels are decreased in height for biasing regions being revisited, thus leading to better convergence of the free energy surface \cite{PRL2008WTMetaD}. The WTmetaD parameters are reported in Table~\ref{tab:metad_parameter}. The height of Gaussian deposition, $\omega$, is selected to 2 $k_{B}T$ at 300 K and the bias factor, $\gamma$, is set to be 100, in order to overcome high energy barrier associated with nucleation problems \cite{salvalaglio2015molecular, Salvalaglio2015Faraday}. The width of the Gaussian, $\sigma$, is set to the thermal fluctuation of corresponding approximate reaction coordinate estimated from a short ($\sim$10 ns) unbiased MD run.

After classifying frames in the biased simulation into either the liquid state or a particular crystal polymorph, the free energy difference between two states $a$ and $b$ at temperature $T$ can be calculated as follows:
\begin{align}
    \Delta G_{a \rightarrow b} = G_b - G_a = k_BT\log{\frac{P_a}{P_b}} 
    \label{eq:FreeEDiff}
\end{align}
where $k_B$ is Boltzmann's constant and $P_a$, $P_b$ denote the Boltzmann probabilities of states $a$, $b$ obtained from reweighting the WTmetaD simulations \cite{tiwary2015jpcb}.

\section*{Results and Discussion}
\label{sec:results}

We performed classical all-atom MD simulations for all systems using GROMACS version-2022.2 \cite{abraham2015gromacs} patched with PLUMED 2.6.1 \cite{plumed2,plumed2019nature} to perform the WTmetaD. Further simulation details are provided in \hyperref[sec:methods]{\textbf{Materials and Methods}}. Below we provide detailed results first for urea followed by glycine.

\setcounter{subsection}{0}
\subsection*{Urea}
\label{urea}

\subsubsection*{Urea Nucleation}
Urea is commonly used as fertilizer and as nitrogen feedstock for organic synthesis. Due to its industrial importance, urea has been studied extensively and several crystalline structures have been synthesized and reported. These include polymorphs named I (space group: $P\bar{4}2_1m$), III ($P2_12_12_1$), IV ($P2_12_12$) and V($Pmcn$), where polymorph I corresponds to the typical isoform stabilized at ambient conditions and the rest are high-pressure, high-temperature products \cite{Weber2002urea,Lamelas2005ureaXray,kat2009urea,donnelly2015urea,bini2017urea,roszak2017urea,safari2021highpressure}. Moreover, an ever-enriched dictionary of polymorphism in urea has been established with computational methods using classical and quantum approaches \cite{giberti2015insight, piaggi2018predicting, Salvalaglio2012uncovering, salvalaglio2015molecular, Salvalaglio2015Faraday, Mandal2017nucleationurea, francia2020systematic, Cheng2017ureaPES}. We adopt the same notations as in Ref.~\onlinecite{piaggi2018predicting}, where two more polymorphs of urea A ($Pnma$) and B($P\bar{1}$) are introduced. It remains unknown if these two crystal structures A and B are artifacts due to use of the generalized Amber force field (GAFF) \cite{amber_gaff} or simply experimentally yet to be observed. Despite the orientational difference in the lattice cell, hydrogen bonding pattern in both forms A and B (all in type III) has been shown to be different from experimental forms (type I and type II) \cite{Cheng2017ureaPES,francia2020systematic}.

Chronologically, we start with a preliminary round of well-tempered metadynamics to facilitate state-to-state transitions along trial reaction coordinates. For this, we biased the two averages of the two feature angles (Fig.~\ref{fig:ang-def} (a)), $\bar\theta_1$ and $\bar\theta_2$ as they have been considered as OPs in distinguishing crystal structures of urea from each other \cite{zou2021sgoopurea}. Several transitions to and from the solid states of interest are observed in the WTmetaD simulations (data not shown) biasing this two-dimensional coordinate (4 independent runs of 400 ns each, totalling $\sim$1.5 $\mu$s) and trained as input data for SPIB along with other candidate OPs. 

\subsubsection*{Urea Polymorphism}
From this preliminary WTmetaD simulation, we performed SPIB to learn a 2-d latent representation for the reaction coordinate (denoted as $z_1$ and $z_2$) as linear combinations of OPs discussed in \hyperref[sec:methods]{\textbf{Materials and Methods}}. We then perform a second round of WTmetaD biasing along this SPIB learned 2-d RC consisting of 4 independent runs of 500 ns each.  The time series for different OPs and the 2-d RC components are shown in Fig.~\ref{fig:urea-sum-md}(a) in black and red, respectively. This unequivocally shows the presence of numerous back and forth state-to-state transitions \revise{(see SI for detailed sampling efficiency comparison)}, a hallmark of good enhanced sampling \cite{valsson2016enhancing}. In particular, three crystal structures, namely polymorphs I, A, and B, are visited during simulations. The corresponding snapshots for these from one are shown in Fig.~\ref{fig:urea-sum-md}(b) along with that of an amorphous bulk phase. The evidence for the formation of these polymorphs can be found in the time series of intermolecular angle OPs (top left two rows in Fig.~\ref{fig:urea-sum-md}(a)), as either one or both of these OPs drop, indicating formation of an ordered phase based on their definitions. \revise{Comparing to the known polymorphs introduced above, structures from experimental products III, IV, V \cite{bini2017urea} and zero-temperature prediction C \cite{Cheng2017ureaPES} are missing. It is possible that the former high-pressure structures are less stable as the existence of mediating solvent molecules, and the latter isoform, could potentially be unstabilized by addition of entropic effects under finite temperature. As these solid states have not been found in previous studies using MD simulations \cite{salvalaglio2015molecular,giberti2015insight} (except form IV being observed in nucleation from the melt \cite{Piaggi2017entropy}), we therefore believe there is remaining scope here for exploration of the configuration space in future studies from other perspectives such as force-field refinement.}

\begin{figure}[t]
  \centering
  \includegraphics[width=1.0\columnwidth]{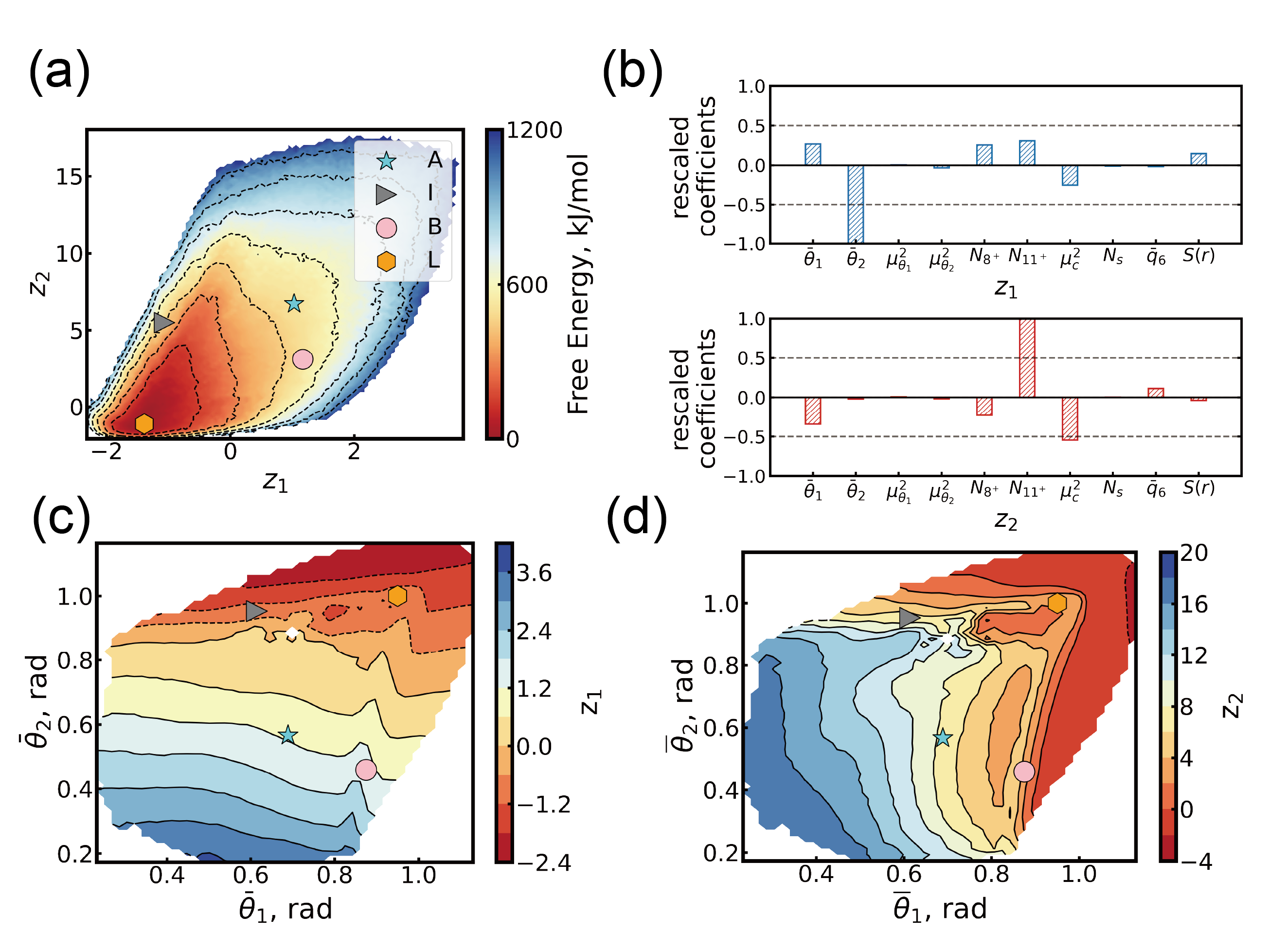}
  \caption
  {(a) Reweighted free energy surface in kJ/mol along SPIB-learned reaction coordinates with landmarks for states of interest (solid states A, I, B, and liquid state L). (b) Rescaled coefficients for each OP in construction of SPIB-learned RCs. The coefficients are rescaled with respect to the corresponding fluctuations of components in WTmetaD simulations (see \hyperref[sec:methods]{\textbf{Materials and Methods}} for details). The importance of the urea molecules relative orientation can be seen from the first reaction coordinate while the second coordinate shows the relevance of clustering. (c) $z_1$-projection onto ($\bar\theta_1$, $\bar\theta_2$) space; and d) $z_2$-projection onto ($\bar\theta_1$, $\bar\theta_2$) space. The markers asterisks (cyan), triangles (grey), circles (pink) and hexagons (orange) indicate crystalline structures A, I, B and liquid phase L, respectively. 
  }
  \label{fig:urea-themo-ana}
\end{figure}

In Fig.~\ref{fig:urea-themo-ana}(a), the reweighted free energy surface along the two-dimensional latent representation $z_1$ and $z_2$ is plotted \cite{tiwary2015jpcb}. The states of interest, one liquid and three solid phases, have been labeled on this surface.
\revise{However, no minima are observed corresponding to regions marked as polymorph states of urea, which could be caused by multiple factors. One of the main reasons suggested is the effect of finite size, which can be corrected analytically using CNT or increasing the size of the simulation box \cite{salvalaglio2015molecular, Salvalaglio2015Faraday}. However, in this work, we emphasize the construction of ML RCs that explore the configuration space reported in previous studies rather than conveying detailed, more quantitative, investigations on thermodynamic and kinetic properties, which are left for future studies.} 

In Fig.~\ref{fig:urea-themo-ana}(b), we provide the coefficients for different OPs, indicating how much they contribute to the RC (see \hyperref[sec:methods]{\textbf{Materials and Methods}} for details on calculating these coefficients). 
From these bar plots, it can be seen that several OPs are weighted heavily in the RC, comprising attributes of intermolecular angle and coordination number. Referring to the definition of these two categories of OPs, the coordination number is a simple but useful OP in the identification of phases, suggestive of the CNT formalism where only size of the nucleating cluster is sufficient. However Fig.~\ref{fig:urea-themo-ana}(b) clearly shows that the coordination number in such complicated molecular systems is not enough for capturing the slow degrees of freedom for crystallization. In addition, our RC indicates that the orientation of feature vectors needs to be taken into account \revise{as these vectors are significant when classifying polymorphic urea \cite{francia2020systematic}}. 

To support the above rationale, we project the SPIB-learned RCs $z_1$ and $z_2$ used to bias onto orientational informative OP space ($\bar\theta_1$, $\bar\theta_2$) in Fig.~\ref{fig:urea-themo-ana}(c) and (d) respectively. Here it can be seen that the driving forces behind the two representations are clearly different. $z_1$ specifically enforces the transitions from initial liquid state (orange hexagon) to states with lower $\bar\theta_2$ values (which are form A (cyan star) and form B (pink circle)) and it shows strong correlation to $\bar\theta_2$ OP itself. On the other hand, $z_2$ pushes the system away from liquid state to all potential solid phases. 


\subsubsection*{Comparison with previous studies}
\begin{figure}
  \centering
   \includegraphics[width=8cm]{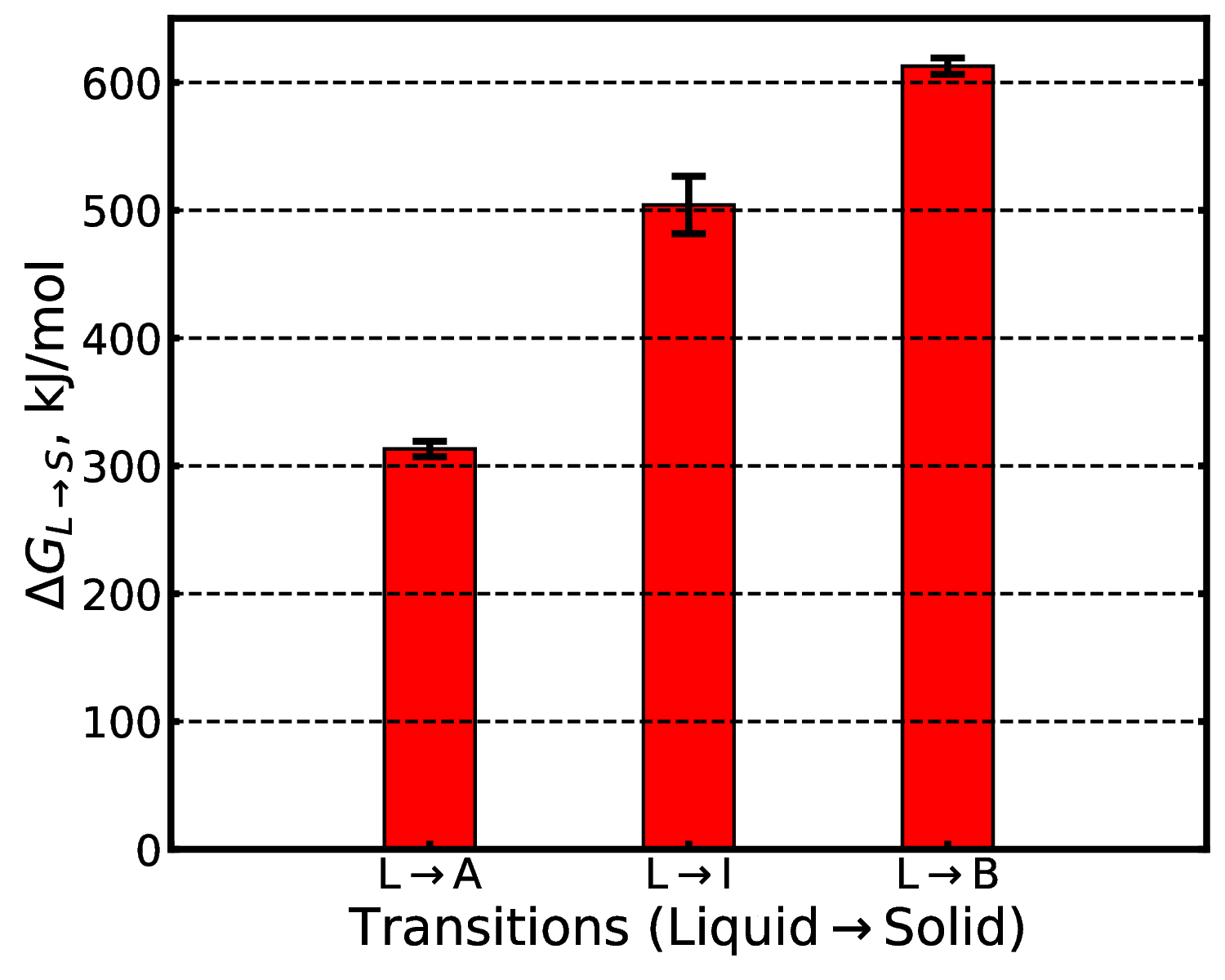}
  \caption
  {Free energy difference in kJ/mol between different \revise{solvated} polymorphs with respect to liquid state as calculated from WTmetaD simulations. This indicates that form A is more stable than form I and form B is the least stable among the three solid phases \revise{under solvated condition}, comparatively. The error bar is computed over four independent product runs. 
  }
  \label{fig:urea-diffG}
\end{figure}

For the purpose of accurate classification of crystal structures for further analyses, the local crystallinity (SMAC) OPs are used (see SI for full expressions) \cite{giberti2015insight, Salvalaglio2012uncovering, salvalaglio2015molecular, Salvalaglio2015Faraday}. In this way, polymorphs can be screened precisely by properly positioning the centers of various switching functions. The size of the nucleus can then be computed using graph theoretic algorithms \cite{Tribello2017analyzing}. After this classification, the free energy corresponding to each phase can be evaluated, giving the free energy difference between two arbitrary states $a$ and $b$ at temperature $T$ as described in~\hyperref[sec:metad]{\textbf{Estimating free energies with Metadynamics}}. Fig.~\ref{fig:urea-diffG} shows the free energy differences of different forms of \revise{solvated, crystalline} urea with respect to the initial \revise{solvated} liquid phase obtained after reweighting the biased WTmetaD simulations \cite{tiwary2015jpcb}. 

\revise{Under the consideration of the existing finite-size effect, we draw only qualitative analyses on these visited metastable states in terms of the relative stability of each crystalline structure in aqueous solution.} As can be seen from this figure (Fig.~\ref{fig:urea-diffG}), our simulations suggest that form A is the most stable polymorph \revise{in the given water model}, followed by form I and finally form B.  \revise{A similar relationship between forms A and I in aqueous solution has been established and studied in detail in previous work: whereas a small nucleus favors crystal structure A, interconversions between A and I occur when cluster grows larger. In particular, Ref.~\onlinecite{salvalaglio2015molecular} suggests a barrierless A-to-I transition occurs as clusters larger than $\sim$50 molecules when applying corrections to leverage finite-size effect, and Ref.~\onlinecite{Mandal2017nucleationurea} claims such a transition occur at a cluster size of $\sim$530 molecules associated to a $\sim$125 $k_BT$ energy barrier when performing a seeding method by inserting pre-built crystals into large simulation boxes.} 

To the best of our knowledge, metastable state B has never been synthesized in aqueous solution of urea, and its very high free energy relative to the other forms may explain why such a structure is difficult to sample. We can explain why we were able to sample this metastable polymorph in our simulations and also in Ref.~\onlinecite{piaggi2018predicting}. Structurally, form B has a completely different orientation among other polymorphs in which its dipole moment in the direction of C-O vector is no longer parallel or anti-parallel to its neighbors. This is likely a result of entropic contributions \cite{piaggi2018predicting}. By having entropy-related OPs in our dictionary of OPs that comprise the RC, we are able to accelerate the substantial entropic degrees of freedom together with other more energy dominated modes, and therefore both enthalpic- and entropic-favored crystalline structures of urea are formed \revise{with enhanced sampling method biasing along machine learned RC}. \revise{We then demonstrate the ability of the SPIB approach on the nucleation of a more complex molecule, glycine, in aqueous solution.}

\subsection*{Glycine}
\label{sec:glycine}
\setcounter{subsubsection}{0}
\subsubsection*{Glycine Nucleation}
\label{subsec:glycinenucleation}
Glycine is the simplest amino acid, with its R-group consisting of a single hydrogen atom \cite{Lehninger2008}. Glycine has multiple physiological functions: it serves as a precursor to more complex biomolecules such as heme, purines, creatine, and more complex amino acids; glycine is also an inhibitory neurotransmitter \cite{Lynch2004} and is involved in cytoprotection and immune system function \cite{Razak2017}. Glycine forms three well-known polymorphs from solution called the $\alpha$ (space group: P2$_1$/n), $\beta$ (P2$_1$), and $\gamma$ (P3$_1$) polymorphs; thermodynamically, $\gamma$-glycine is the the most stable, but $\alpha$-glycine tends to be the form that crystallizes from solution at neutral pH and ambient atmospheric pressure \cite{Bushuev2017, Boldyreva2003, Yani2012, Boldyreva2021}. Several other high-pressure polymorphs have been discovered \cite{Boldyreva2021, Bull2017}, but they are not typically observed at atmospheric pressures or temperatures. For performing the WTmetaD \cite{PRL2008WTMetaD}  simulations of glycine, we use a similar OP library as urea, with one major exception: the pair entropy $S(r)$ is swapped for the orientational pair entropy $S(r,\theta)$ \cite{piaggi2018predicting}. The angles used for calculating $S(r,\theta)$ are those given in Fig.~\ref{fig:ang-def}(b). For glycine, the first intramolecular vector is defined as that from the N-terminal nitrogen to the C-terminal carbon atom; the second intramolecular vector starts at the alpha-carbon C$_{\alpha}$, and ends at the R-group hydrogen atom H$_{\alpha}$.

In contrast to urea, we find it necessary to include $S(r, \theta)$ in the OP library because, without it, reversible nucleation events are not observed (data not shown). We postulate that this increased importance in the case of glycine is due simply to $S(r,\theta)$ being a two-body term and, hence, a first-order correction to the excess entropy of a fluid \cite{Prestipino2004, Baranyai1989} compared to the entropy of the equivalent ideal gas. Since glycine is more massive and zwitterionic, as well as possessing a stronger dipole moment, it is less ideal in the liquid phase compared to urea and, therefore, it is to be expected that it will possess more excess entropy compared to urea, making it necessary to place $S(r, \theta)$ and not just  $S(r)$ in the OP library.

For the preliminary round of WTmetaD, we bias along $S(r,\theta_1)$ and $S(r,\theta_2)$ since these OPs were found to give the best sampling from the OP library. This 400ns preliminary simulation and its SPIB analysis is described in detail in the SI. As with urea, the trajectories of the OPs from this preliminary simulation are used as input data for training SPIB. \revise{Also in the SI are projections of the metadynamics bias onto the ($\bar{\theta}_1$, $\bar{\theta}_2$) space, where it can be seen that using the SPIB RCs greatly enhances the amount of bias deposited, and hence the amount of sampling, of regions of ($\bar{\theta}_1$, $\bar{\theta}_2$) space where the glycine polymorphs of interest reside. This result provides support for using the SPIB RCs as the biasing coordinates over the hand-picked orientational entropies.}

\subsubsection*{Glycine Polymorphism}
\label{subsec:glcyinepolymorphism}
Here we are interested in the $\alpha$-, $\beta$-, and $\gamma$-glycine polymorphs observed at ambient temperature and pressure \cite{Boldyreva2021}. A necessary but not sufficient method \cite{Iitaka1961} to differentiate the three polymorphs is through the collective orientation of the C-N vectors (Fig.~\ref{fig:ang-def}(b)). For $\alpha$-glycine, C-N vectors alternative in layers of two in a `positive' orientation (C-N) followed by a `negative' orientation (N-C) repeating. For $\beta$-glycine, positive and negative layers alternate one at a time. In $\gamma$-glycine, all layers of the crystal possess the same orientation, either all positive or all negative. Using only the C-N axis orientations to classify polymorphs, we find a relative stability rank order of form-$\gamma >$ form-$\beta \ge$ form-$\alpha$, which recovers form-$\gamma$ as the most stable form in agreement with experiments of bulk glycine \cite{Boldyreva2003,Boldyreva2021}.

While this classification approach correctly accounts for the large-scale orientation of the crystal, it ignores the local orientation, such as hydrogen bond patterns and relative angles and distances between layers of the crystal \cite{Marom2013, Boldyreva2021}. This coarse procedure for identifying polymorphs is selected over the more rigorous one given in Ref.~\onlinecite{duff2011polymorph}, which has not be tested for use in nucleation from solution or long timescale simulations, where there can be significant perturbations to the crystal structure.
Based on these considerations, we have chosen to characterize the polymorphs using the relative orientation of monomers along the axis parallel to the C-N vector only. 
Ignoring local structure, this protocol will introduce some contamination of the polymorph classification. Furthermore, the crystal structures of glycine are much more similar to each other compared to urea, making the SMAC protocol used for urea ineffective for glycine, with details for why this is the case given in the SI. Since this study is, to our knowledge, the first to perform enhanced sampling simulations of glycine to study nucleation without seeding in water alone at or below the saturation concentration, our reported numbers at the very least have qualitative significance in being able to correctly rank order glycine polymorph stability in pure aqueous solution. 

\begin{figure*}[t] 
\center
\includegraphics[width=1.0\linewidth]{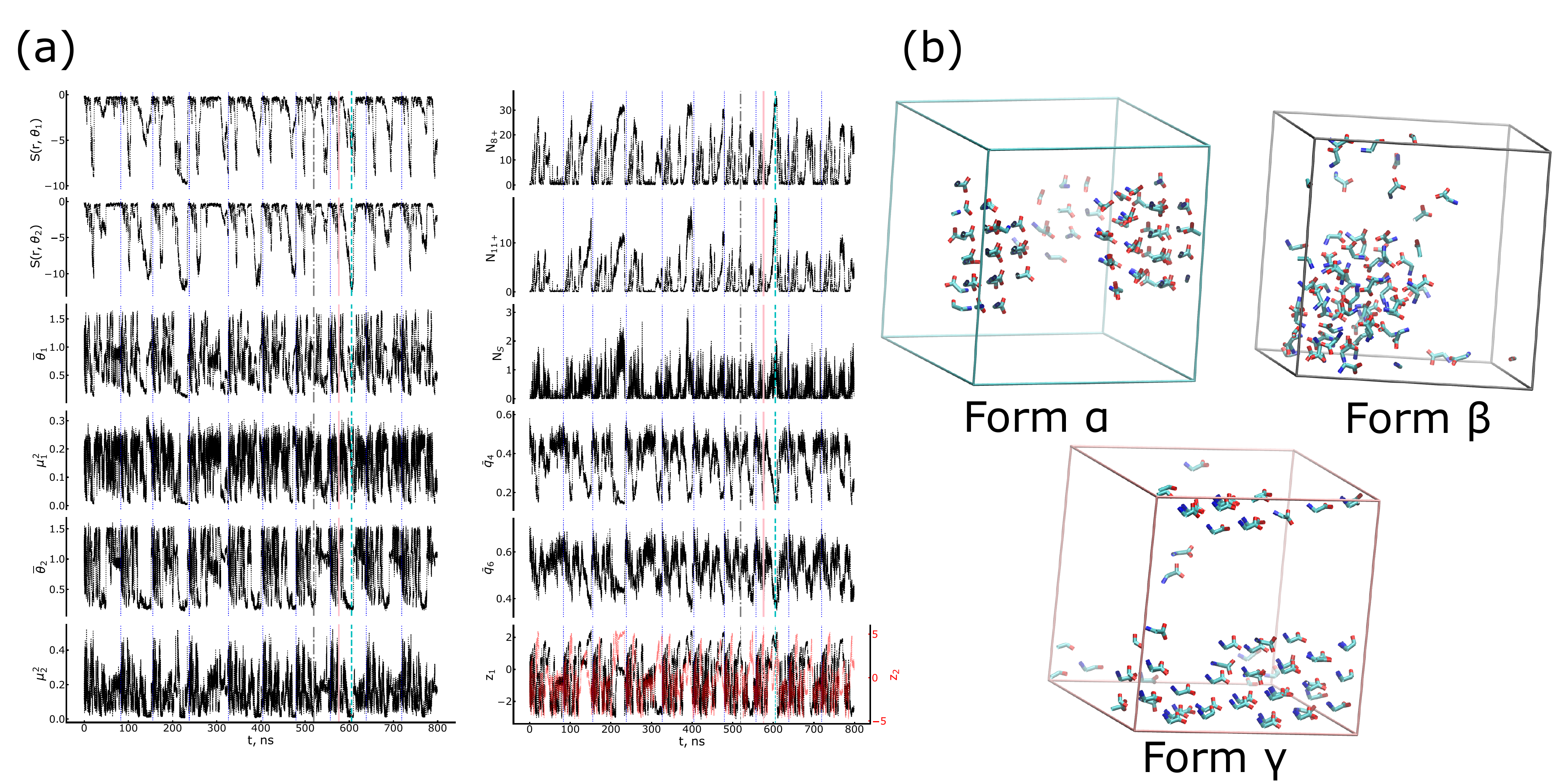}
\caption{Sampling glycine polymorphs with WTmetaD simulations biasing the SPIB-learned 2-dimensional reaction coordinate. (a) shows the time series of the OPs (black) and SPIB RCs (black, $z_1$; red, $z_2$) from the ten replicate simulations of glycine, with each trajectory demarcated by the dotted blue lines. Vertical cyan dashed, gray dashed-dot, and pink solid lines indicate representative transitions to form $\alpha$, form $\beta$, and form $\gamma$ from the liquid phase. (b) shows snapshots of the polymorph structures sampled during trajectories shown in (a); color of the box framing each snapshot corresponds to the color of the vertical time slice in (a). Hydrogen atoms on glycine and water molecules are removed for clarity; snapshots are rendered using VMD.}
\label{fig:snapshots_glycine}
\end{figure*}

Fig.~\ref{fig:snapshots_glycine}a shows \revise{the trajectory of the OPs used in the SPIB analysis for glycine, and, in the final panel, the trajectories of the two SPIB RCs; all trajectories come from the final set of simulations biased along the discovered linear SPIB RCs. These trajectories clearly show many transitions between the solvated, crystalline states and the liquid state of glycine; a comparison of the sampling efficiency of this set of trajectories biased along the SPIB RCs and the hand-picked orientational entropy CVs is given in the SI for both glycine and urea.} 

Fig.~\ref{fig:snapshots_glycine} shows the $\alpha$-glycine, $\beta$-glycine  and $\gamma$-glycine polymorphs extracted from the SPIB-biased trajectories. For all three snapshots, the appropriate orientations for each polymorph along the monomer C-N axis should be apparent, although, given the size of the system (72 glycine molecules), finite volume and surface area effects are likely to be significant due to the large surface area to volume ratio of the clusters. We also expect solvent-induced perturbations compared to the respective crystal structures \cite{Iitaka1960, Dawson2005} from the melt. 

\begin{figure*}[t] 
\center
\includegraphics[width=0.9\linewidth]{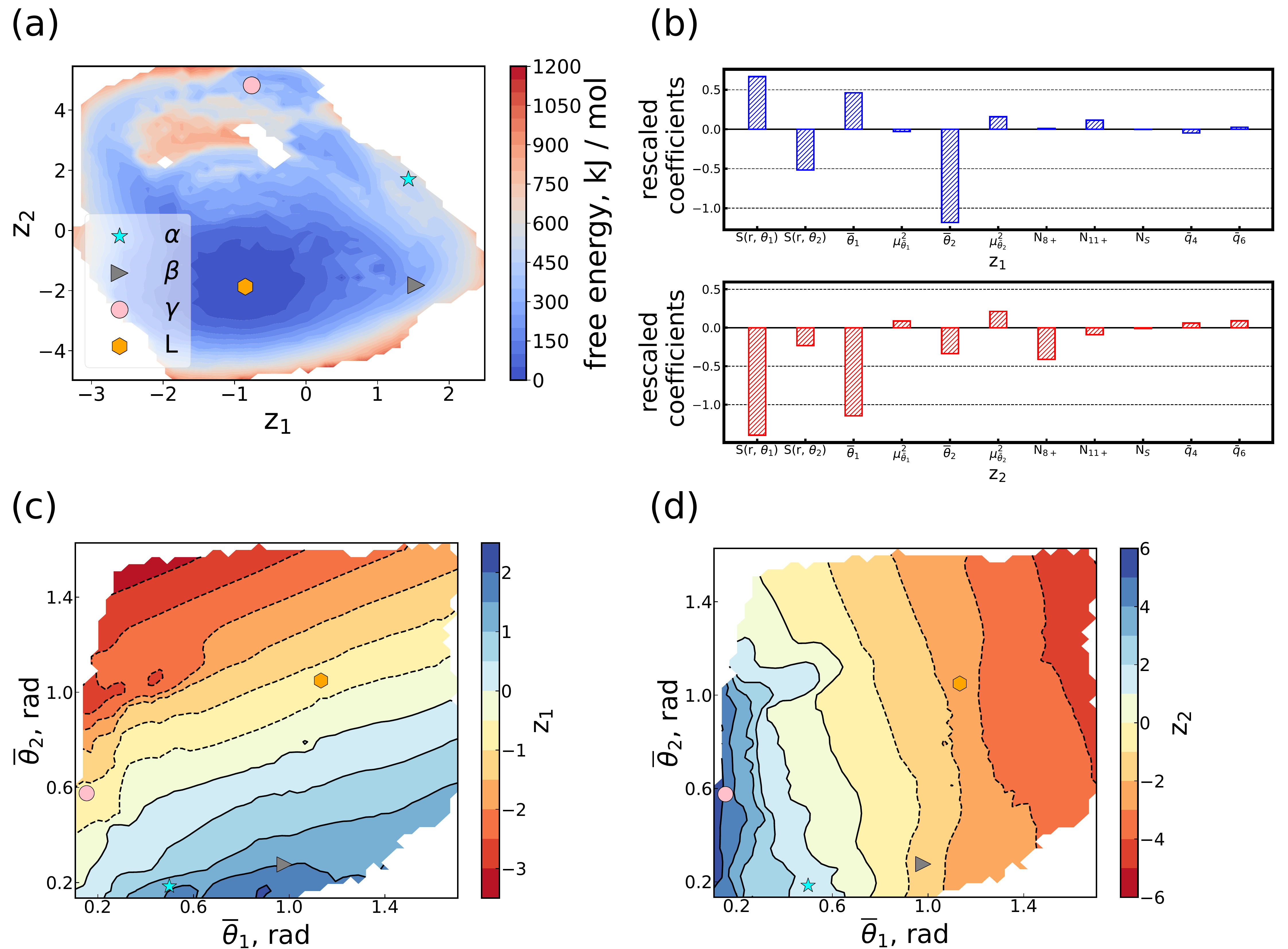}
\caption{(a) Reweighted free energy surface in kJ/mol in the two-dimensional space of the linear SPIB RCs, which were used as the metadynamics biasing variables. Markers show the locations of the putative glycine polymorphs sampled during the simulation: liquid, orange hexagon; $\alpha$, cyan star; $\beta$, grey sideways triangle; and $\gamma$, pink circle. (b) SPIB coefficients from the linear model, reweighted by the standard deviation of the input variable. For the first SPIB RC (top panel, blue bars), the most important input variables is the trajectory of the average second intermolcular angle $\bar{\theta}_2$, and the orientational entropy S(r, $\theta_1$). For the second SPIB RC (bottom panel, red bars), the two most important input variables are $\bar{\theta}_1$ and its orientational entropy S(r, $\theta_1$). (c) The first SPIB RC z$_1$ projected into the ($\bar{\theta}_1, \bar{\theta}_2$) subspace. Colored symbols show the locations of the putative glycine polymorphs, with the coloring identical to that in panel (a). (d) is the same as (c), except the projection is of the second SPIB RC, z$_2$.}
\label{fig:spib_glycine_analysis}
\end{figure*}

An analysis of the SPIB RCs used to bias the nucleation simulation of glycine is given in Fig.~\ref{fig:spib_glycine_analysis}. In Fig.~\ref{fig:spib_glycine_analysis}(a), the reweighted free energy surface along the 2-d linear SPIB latent space used as the biasing variables in metadynamics is shown with the locations of the putative $\alpha-$, $\beta$-, and $\gamma$-glycine polymorphs labeled with colored stars; for reference, the liquid (isotropic) state is also labeled with a magenta star. Fig.~\ref{fig:spib_glycine_analysis}(a) shows that, while none of the glycine polymorphs are located in free energy minima they are neither at free energy maxima, which indicates that the SPIB RCs are doing a reasonable job sampling the crystal polymorphs.

Fig.~\ref{fig:spib_glycine_analysis}(b) shows the rescaled contribution of each input variable to the final two-dimensional SPIB model. For both RCs, the orientational entropies are weighted heavily in-line with the physical intuition explained in~\hyperref[subsec:glycinenucleation]{\textbf{Glycine nucleation}} that higher-order corrections to the entropy of the zwitterionic glycine will be more important than an uncharged molecule such as urea. Also highly weighted are the two angular coordinates, $\bar{\theta}_1$ and $\bar{\theta}_2$, which justifies performing the polymorph labeling and analysis in that subspace.

\begin{figure}[t] 
\center
\includegraphics[width=0.8\linewidth]{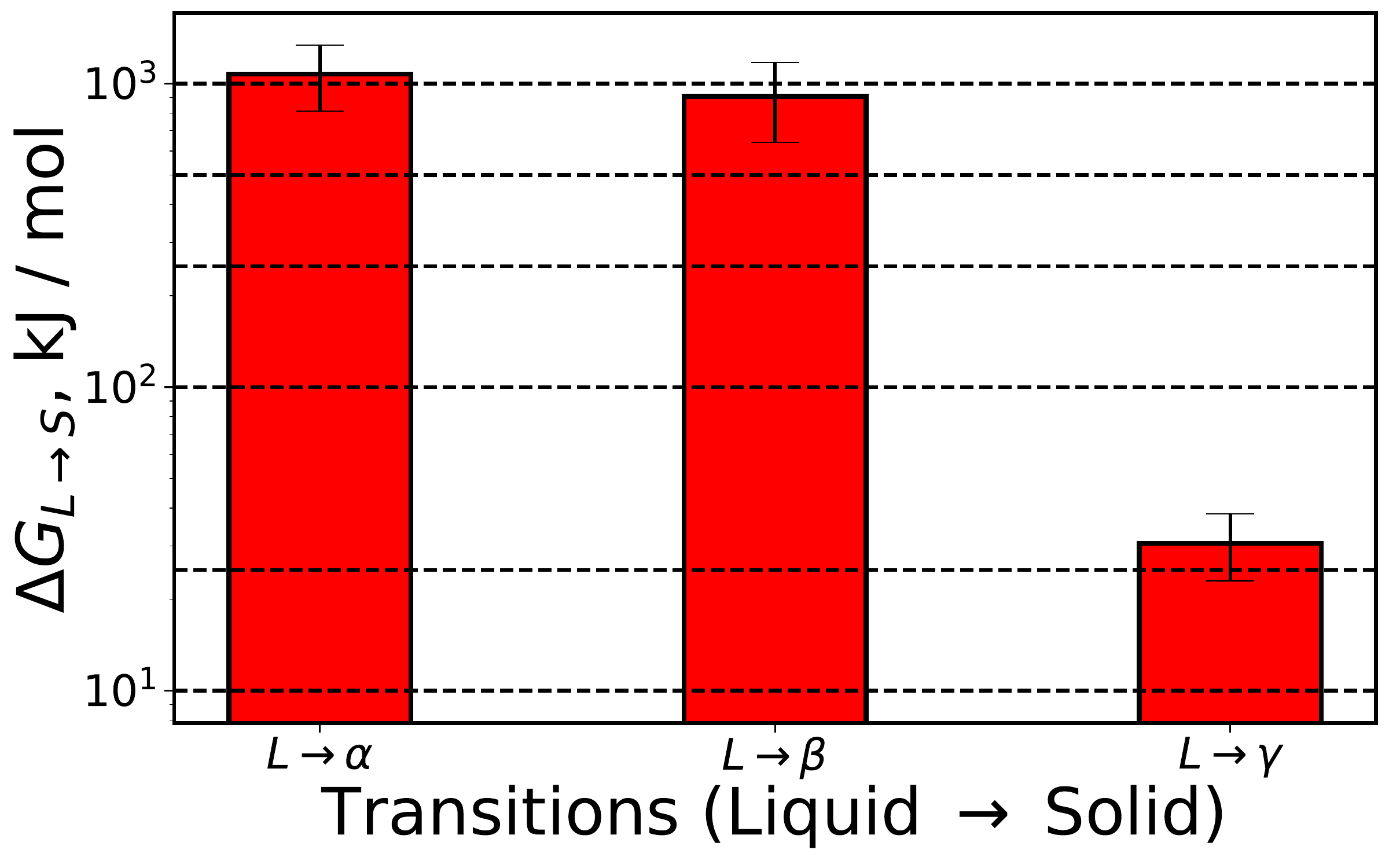}
\caption{
Free energy difference in kJ/mol between different \revise{solvated} polymorphs with respect to liquid state as calculated from WTmetaD simulations. This result indicates that form $\gamma$ is more stable than form $\beta$, which is slightly (though not significantly) more stable than form $\alpha$ \revise{when solvated with the solvated liquid state as a common reference}. Error bars are one standard error of the mean over the ten replicate simulations.}
\label{fig:dG_glycine}
\end{figure}

The SPIB RCs are further interpreted by projecting them onto the ($\bar{\theta}_1, \bar{\theta}_2$) subspace in Figs.~\ref{fig:spib_glycine_analysis}(c), (d). The first SPIB RC, $z_1$, corresponds to nucleation from the liquid to the three common polymorphs labeled on the surface. Examining the contours more closely, it specifically describes transformation from the liquid and $\gamma$-glycine states, which are the two most populated states observed in our simulations, to the $\alpha$- and $\beta$-glycine polymorphs. There is also a strong correlation of $\bar{\theta}_1$ with z$_1$, as expected from Fig.~\ref{fig:spib_glycine_analysis}(b). The second SPIB RC, z$_2$, corresponds to transitions from the liquid state to $\gamma$-glycine via the $\alpha$ and $\beta$ polymorphs. Interestingly, this mechanism of transition from the liquid state to the most stable crystal polymorph (form $\gamma$) via the less stable crystal polymorphs (forms $\alpha$ and $\beta$, Figure \ref{fig:dG_glycine}) is consistent with the Ostwald step rule \cite{vansanten1984ostwaldsteprule} and is evidence for two-step nucleation in aqueous glycine, which already has been observed experimentally\cite{Urquidi2022}. This RC also correlates strongly with transitions from low-to-high values of S(r, $\theta_1$) (data shown in the SI), again as expected from Fig.~\ref{fig:spib_glycine_analysis}(b). 

In summary, we find the linear RCs discovered by SPIB are able to 1) separate the common glycine polymorphs in the two-dimensional RC space and 2) yield a reasonable and interpretable set of reaction coordinates for describing nucleation of glycine from aqueous solution.

\subsubsection*{Comparison with \revise{previous studies}}
\revise{While the SMAC OP is used to classify urea polymorphs, we find that the g(r, $\theta$) distributions for the polymorphs of glycine to be too similar for SMAC to be effective in distinguishing polymorphs and instead using the classification method outlined in~\hyperref[subsec:glcyinepolymorphism]{\textbf{Glcyine polymorphism}}.We also propose a second method based on Voronoi tesselation of the ($\bar{\theta}_1$, $\bar{\theta}_2$) space around the `landmark' polymorph and liquid structures shown in Figure \ref{fig:spib_glycine_analysis} in the SI.} 

To our knowledge, no \revise{explicit} free energy differences between polymorphs for crystallization from aqueous solution for glycine have been reported in the literature. Solubility experiments in pure water \revise{and water-organic solvent-antisolvent mixtures of varying composition} have been reported in Ref.~\onlinecite{Bouchard2007}, \revise{where the authors show that, at each solvent composition (ranging from 80\% water/20\% methanol to 0\% water/100\% methanol) the solubility order is $\beta$ $>$ $\alpha$ $>$ $\gamma$. Using the argument that the solubility of the crystal in solvent decreases as the free energy barrier increases (since the solubility is roughly the ratio of the concentration of the solvated liquid form to the solvated, crystal form) \cite{Parks2017,salvalaglio2015molecular}, this solubility order implies a \revise{relative} stability order \revise{with reference to the solvated liquid state} of  $\gamma$ $>$ $\alpha$ $>$ $\beta$, which is in rough agreement with the relative stability we find, namely $\gamma$ $>$ $\beta$ $\ge$ $\alpha$.}

\revise{We additionally note that the only study performed to rank the relative stability of glycine polymorphs is Ref.~\onlinecite{Parks2017}, where the authors find that the $\gamma$-form of glycine is the least kinetically accessible and the $\beta$-form the most kinetically accessible, at the nanoscale, using calculations based on seeded cluster MD simulations and CNT. Since we use a vastly different setup (classical MD coupled with WTmetaD and machine learning versus seeded MD and CNT), we are not concerned about disagreements between polymorph accessibility in our study and the one performed in Ref.~\onlinecite{Parks2017}. However, we concede that our glycine clusters are well below the critical nucleus size predicted by CNT and thus may be plagued by finite size effects \cite{Honeycutt1984, Swope1990}, especially considering that many times the $\gamma$ polymorph is observed to form in the simulations, the crystal spans the simulation box along one dimension, leading to an infinite crystal when periodic boundary conditions are taken into account (see Fig.~\ref{fig:snapshots_glycine}); this effect could be a primary reason we observe that polymorph to be sampled the most frequently. Additional study is needed to determine the extent to which finite size effects affect the relative stabilities of glycine polymorphs reported here. }

\revise{We conclude this section by noting that while glycine is a common molecule for experimental study of nucleation and crystal polymorphism, it is also a notoriously difficult system to obtain consistent crystallization results experimentally \cite{Boldyreva2003,Boldyreva2021}. For example, although most experiments report that the $\alpha$ polymorph is the primary crystal product extracted from solution \cite{Yani2012, Urquidi2022, Cheong2010,Broadhurst2020, Little2015}, the $\gamma$ form is observed to form spontaneously and nearly exclusively from solution given the appropriate conditions \cite{He2006, Hughes2007, Hughes2009, Boldyreva2021, Little2015, Aber2005}. Furthermore, the timescales of most experimental studies of glycine nucleation tend to be orders of magnitude longer (hours to days) than the MD simulations studied here, making comparisons to experimental results difficult (e.g. the kinetic product seen in those experiments could be the thermodynamics product observed in this study). However, regardless of any quantitative comparison to experiments, we are encouraged by the results presented, as we are able to sample the three well-known experimental polymorphs of glycine using the SPIB approach. }
\section*{Conclusion}
\label{sec:conclusion}

Although nucleation is a frequent event on the experimental and biological timescales, with currently accessible compute resources nucleation is a rare event computationally, even when utilizing enhanced sampling techniques \cite{zou2021sgoopurea, giberti2015crystalnucleation,giberti2015insight, karmakar2019constantmu,Giberti2013,Salvalaglio2016argon,tsai2019reaction}. Thus, even for relatively simple systems such as the aqueous urea and glycine systems studied here, sampling polymorph configurations and, more dauntingly, determining relative thermodynamic stabilities of these polymorphs, is difficult computationally. In this work we have demonstrated a more-or-less automatable protocol to obtain back-and-forth sampling of different polymorphs in aqueous solution. Specifically, we have shown that good sampling of polymorphs is possible when classical molecular dynamics simulations of urea and glycine are biased along optimized reaction coordinates found using the machine learning approach State Predictive Information Bottleneck (SPIB) \cite{Wang2021SPIB} which belongs to the RAVE family of methods \cite{wang2019past,wang2020machine}. A linear, two-dimensional approximation to the reaction coordinate extracted from SPIB \cite{Wang2021SPIB} can be used to 1) effectively bias nucleation simulations of aqueous urea and glycine to enhance the sampling of polymorph nucleation, and 2) interpret the SPIB RCs to determine which OPs are the most important drivers of nucleation.

When interpreting the RCs derived from the SPIB analysis, we find that for urea, the intermolecular angle and cluster size OPs are the most important contributors to the RC; for glycine, the intermolecular angles and orientational entropy of these angles are the most important to determining the reaction coordinate. These observations reflect that size informative OPs suggested by CNT are insufficient in describing slow modes towards nucleation of urea and glycine molecules; in addition orientational OPs are needed. Each polymorph is identified using different metrics and then ranked appropriately as per their Boltzmann weights at 300 K \cite{tiwary2015jpcb}. The relative stability \revise{of the solvated crystal polymorphs compared to the solvated liquid state} found by us shows  form-A $>$ form-I $>$ form-B for urea and form-$\gamma >$ form-$\beta \ge$ form-$\alpha$ for glycine. Since we performed the SPIB analysis on the unreweighted energy surface, we cannot yet offer any kinetic insights regarding the nucleation of either organic molecule studied here. In addition we did not consider the question of finite size effects, which would especially impact kinetic properties. This will be done in future work using approaches such as Ref.~\onlinecite{Salvalaglio2016argon}. 
Secondly, our simulations would have been more representative of supersaturation levels of experiments with the use of constant chemical potential ensemble, which is an active area of research for studying nucleation \cite{karmakar2019constantmu}.

To summarize, this work makes several significant contributions to the field of nucleation research and enhanced sampling in general. We have applied deep learning based RC construction methods to arguably the most complex set of systems to date and used them to successfully find reaction coordinates that accelerate the nucleation process of multiple polymorphs for both urea and glycine aqueous solutions. Furthermore, the simulations of glycine reported here are the first, to our knowledge, to utilize metadynamics to induce nucleation of the amino acid in water. Based on these results, we believe using machine learning methods to construct approximate reaction coordinates can provide novel solutions and information regarding non-classical effects in the nucleation of molecules from aqueous media.

\section*{Associated Content}
\label{sec:associatedcontent}
\subsection*{Supporting Information}
The Supporting Information is available free of charge at xxx. It contains detailed further numerical analyses for different systems.


\subsection*{Notes}
The code needed to reproduce the models used in this work is
available at https://github.com/tiwarylab/Driving-and-characterizing-nucleation-of-urea-and-glycine-polymorphs-in-water. 
The input files necessary to reproduce the simulations done in this work are available on PLUMED NEST at \url{https://www.plumed-nest.org/eggs/22/039/}.

\section*{Materials and Methods}
\label{sec:methods}
\subsection*{Simulations setup}
A 5.0 nm $\times$ 5.0 nm $\times$ 5.0 nm simulation box containing 300 urea and 3085 TIP3P \cite{Jorgensen1983tip3p} water molecules is built to mimic the work of Salvalaglio \textit{et al} \cite{salvalaglio2015molecular}. The partial charges of urea were adopted from the Amber03 \cite{amber03} database. For the aqueous simulations of glycine, we follow the simulation protocol established in Ref.~\onlinecite{Bushuev2017mdGlycine}. A simulation box of size 3.6 nm $\times$ 3.6 nm $\times$ 3.6 nm is constructed with 72 glycine molecules and 1200 SPC/E \cite{berendsen1987spce} water molecules with generalized Amber force field (GAFF) \cite{amber_gaff},  and the charges on glycine are those derived using the CNDO methodology in Ref.~\onlinecite{Derissen1977}. This combination of the GAFF and CNDO charges for glycine is utilized because it has been found to accurately reproduce the physical properties of both aqueous and crystalline glycine as well as inducing $\alpha$-glycine crystallization from solution \cite{Cheong2010}. The glycine simulations correspond to a saturated glycine solution (whose molality is experimentally measured to be 3.33 mol/kg) \cite{Bushuev2017mdGlycine}. \revise{However, since we do not explicitly calculate the solubility of glycine with the GAFF parameterization using the SPC/E water model, the solubilty and the saturation level of glycine is unknown during the simulations.}

For both the systems, all MD and metadynamics simulations were performed in the constant molecule number, pressure, temperature (NPT) ensemble with an integration time step of 2 fs. Systems were coupled with a thermostat of velocity rescaling scheme \cite{bussi2007canonical} at 300 K. The pressure was controlled using the Parrinello-Rahman barostat \cite{parrinello1981polymorphic} at 1 bar. The relaxation times to thermostat and barostat were 0.1 ps and 1 ps, respectively, for both system investigated. Particle-mesh Ewald method \cite{pme} was used for the computation of long range electrostatic interaction and hydrogen bonds were constrained with the LINCS algorithm \cite{lincs}. The cutoffs of electrostatic and Van der Waals interaction in real space were selected to be 1.2 nm for glycine and 1.0 nm for urea.

\subsection*{Rescaling Coefficients of SPIB-learned Representations}
As the coefficients learned by SPIB are in dimensions of the inverse of the corresponding OPs and unstandardized as input, proper rescaling of these coefficients is needed for the purpose of evaluating the importance of each OP to the RC. Following the ``betasq'' protocol from Ref.~\onlinecite{Groemping2006}, these rescaled coefficients are calculated as the product of the coefficients learned by SPIB and the fluctuation of the OP individually from the input trajectories (WTmetaD simulations biasing $\bar\theta_1$ and $\bar\theta_2$ OPs).

\section*{Acknowledgments}
\label{sec:acknowledgements}
The authors thank Dr. Pablo M. Piaggi for providing source codes for the  entropy OPs. We also thank Prof. John D. Weeks, Prof. Matteo Salvalaglio, Dr. Yihang Wang, Dr. Ruiyu Wang, Zachary Smith, Luke Evans, Dedi Wang, and Renjie Zhao for discussions and Dr. Ruiyu Wang, Luke Evans for proofreading the manuscript. This research was entirely supported by the U.S. Department of Energy, Office of Science, Basic Energy Sciences, CPIMS Program, under Award DE-SC0021009. We also thank Deepthought2, MARCC and XSEDE \cite{Towns2014} (projects CHE180007P and CHE180027P) for computational resources used in this work. 

\showacknow{}
\bibliography{references}

\end{document}